\documentclass{ws-p9-75x6-50_3}

\usepackage{tfw2}
\usepackage[]{graphicx} 
\begin{document}
\title{On the Fine Structure in the Fireballs Peak of the 1998 Leonids}
\author{COSTANTINO SIGISMONDI}
\address{Yale University, Dept. of Astronomy, 260 Whitney Ave. New Haven, CT 06511 USA;
 Astronomical Observatory of Rome and ICRA, International Center for Relativistic Astrophysics,
Physics Dept., University of Rome``La Sapienza"  P.le A.Moro 2 00185 Rome Italy~~~
e-mail:~~ sigismondi@astro.yale.edu~~~~~sigismondi@icra.it}
\maketitle
{\sl The Leonids meteor shower of November 1998 has shown double activity. An
unexpected shower of fireballs occurred about 16 hours before the
expected maximum of the meteor activity [1]. The activity profile of the
fireballs shower also revealed  fine structure. The 
hypothesis for a tidal origin of such sub-structures is discussed. The close encounter of the
meteoroid stream with the Earth in 1366 and the encounters 
with Saturn (in 1630) and Jupiter (in 1732), are identified as the
cause of the main features of the structure. Such analysis can be applied to the 
incoming 2000 Leonids' display [2].}
\begin{figure}
\resizebox{6.2cm}{3.5cm}{\includegraphics{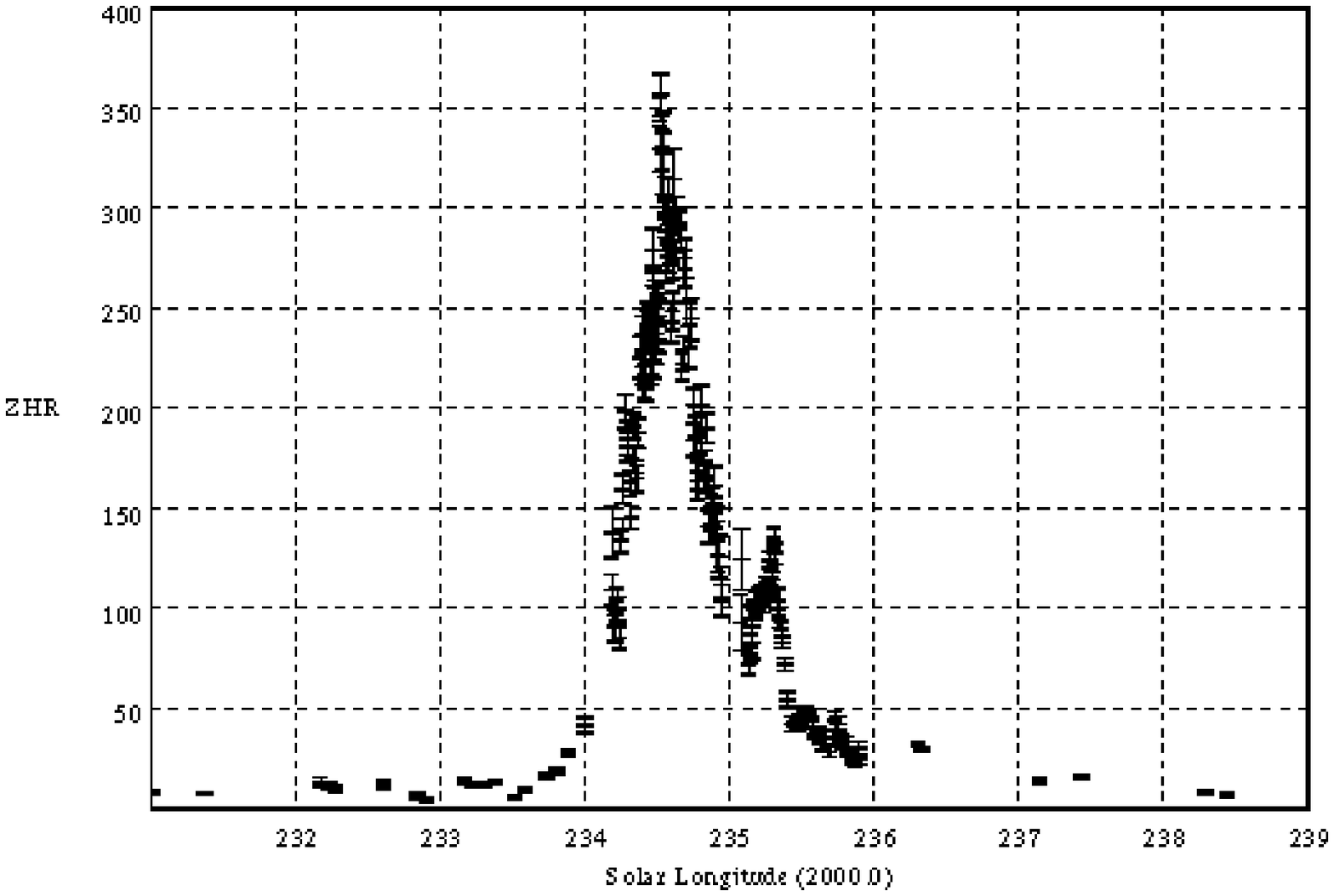}}
\hspace{0.05cm}
\resizebox{6.2cm}{3.5cm}{\includegraphics{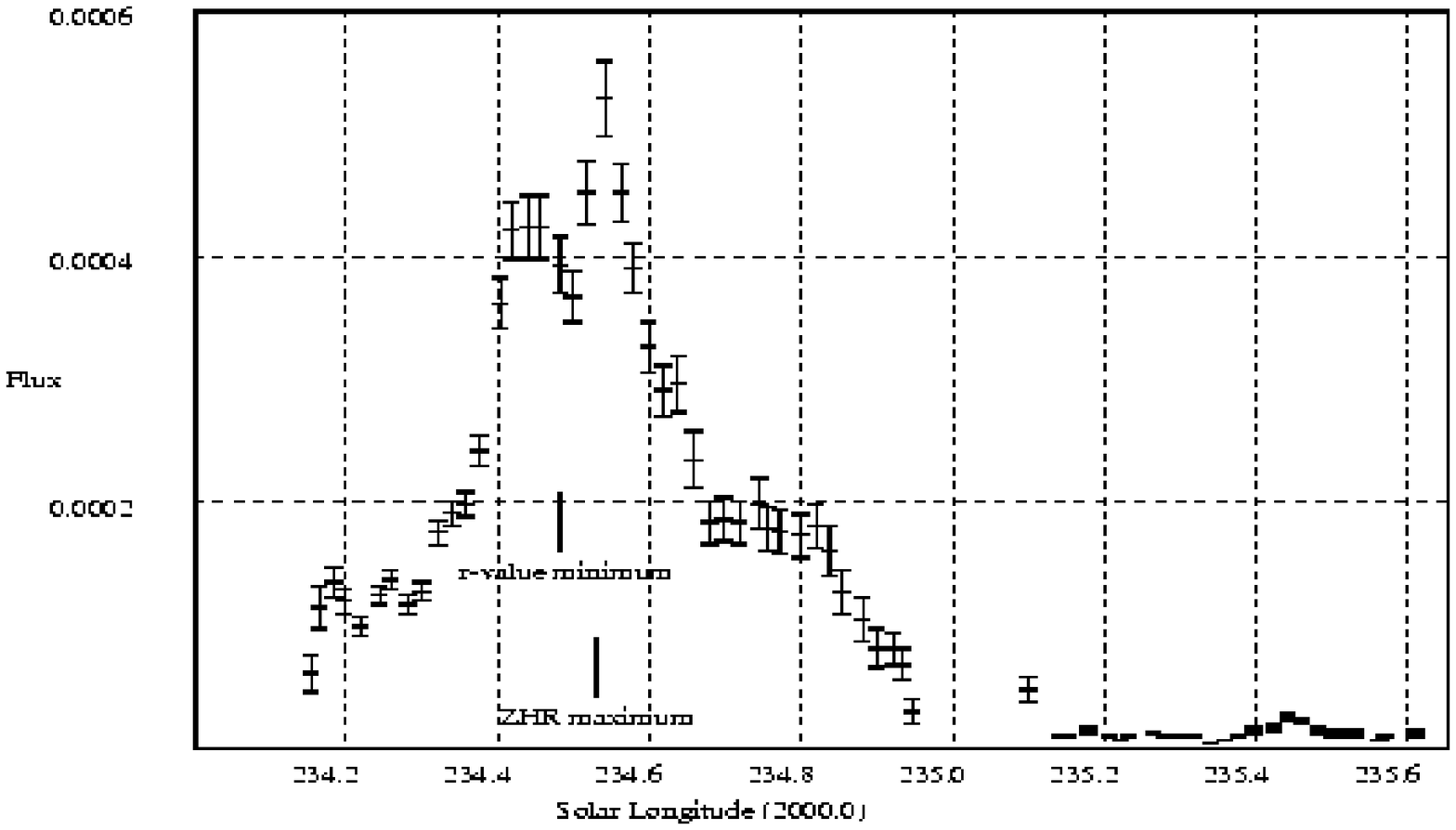}}
\caption{1998 Leonids' activity, data from [1]. \textbf{Left:} The fireballs
peaked at $\lambda_{\odot}\sim234.5$. The `storm' component occurred
at $\lambda_{\odot}\sim235.3$, with the population index $r\sim2$ 
(if $N_m$ is the number of meteors fainter than magnitude 
$m$, $N_{m+1}=N_m \cdot r$, lower values of $r$ mean higher 
$\%$ of bright meteors). The fine structure appeared
at $\lambda_{\odot}\sim234.5$. %\linebreak
\textbf{Right:} The double peak 
in the fireballs flux. 
The number of bright meteors per Km$^2$ versus solar longitude is plotted.  
Between $\lambda_{\odot}=234^o 4$ and $234^o 55$ we find that $r$ was as low as $r\sim1.3$.}
\end{figure}
The meteoroids 
ejected during the 1333 perihelion passage of the parent comet 55P/Tempel-Tuttle (or even before) were 
trapped in the stable resonance zone 5/14 mean-motion with Jupiter [3]. 
In 1998 the Earth crossed this zone 16 hours before crossing the node of the parent comet's 
orbit. 
%Planetary perturbations produce small secular changes of the comet's orbit.
In contrast to the 1866 Leonids storm, 
when a filamentary
structure was claimed, yet not confirmed by
comparing data 
of different astronomers [4], the meteors in 1998 were
all very much brighter than any local limiting magnitude, so they were
easily visible in the city lights of Rome.
The double peaked structure of the fireballs reflects the
previous dynamical history of the cometary debris.
% \linebreak
Supppose that a bimodal debris cloud is initially separated by a distance $\delta$,
then, after the tides, the final separation becomes %\noindent
 $D=v_{rel}\frac{\Delta t_{obs}}{cos\theta}=650000$ Km,  $v_{rel}=71$ Km s$^{-1}$ 
is the velocity of the meteoroids relative to Earth,  and $\Delta t_{obs}=2.6$ hours is the temporal 
separation between the two peaks. Let the separation be perpendicular, ``$\bot$", 
to the comet's path (relative to the Earth) with its orbital plane inclined by $\theta\sim 18^o$
\newpage
\noindent
 with respect to the ecliptic.
Several jets of gas and icy dust occur on the comet's surface. 
The largest meteoroids ($\ge 1$ cm, $\ge 10$ g) orbit tightly around the 
comet's nucleus within $\delta=25000$ Km (a typical {\sl coma} dimension). The Earth's gravitational field 
gives them the escape velocity $v_{esc}\sim 2$ m s$^{-1}$ to reach the resonant zone and the tidal 
impulse $\Delta v_{\bot,\oplus}\sim2$ cm s$^{-1}$  to reach $40\%$ of the current separation D, after $632$ years. 
Saturn and Jupiter power the initial impulse for the remaining $60\%$ of $D$.
Referring to the geometry described in figure 2 we obtain: 
$\Delta v_{\bot,\oplus}=\int_{0}^{\infty}{F_{tidal}dt\over m}\
%,dt\cr
=2\int_{0}^{\infty}\left({GM_{\oplus}\over x^2+b^2}-{GM_{\oplus}\over
x^2+(b+\delta)^2}\right)\,{b\over \sqrt{x^2+b^2}}\,{dx\over v_{rel}}
%\cr
\approx {4GM_{\oplus}(\delta^2+2b\delta)\over 3v_{rel}b^3}\,$
%,\cr}$
%
where $m$ is the mass of an average meteroid, $M_{\oplus}$ the Earth's mass.
The solution of the equation  
$\Delta v_{\bot, \oplus}nT_{orb}\approx 0.4\cdot D$,
with $T_{orb}=33.25$ years, is
a set of pairs $(n,b)$. Varying the integer $n$, we can compare
the corresponding value of $b$ with the values of the geocentric
distance of the comet during previous perihelion passages [5]. 
For $n=19$ (i.e. 1366), the above
equation yields $b$ in agreement with the calculated geocentric distance at that time.
This result agrees with the hypothesis that
the fireball particles were ejected in 1333 [3], and it is obtained
independently from temporal fine structure observations.
The tidal actions of Saturn in
1630 (at 0.34 AU) and Jupiter in 1732 (at 0.83 AU) [5] give impulses
$\Delta v_{\bot} \approx \Delta v_{\bot,\oplus}\sqrt{a}\,{M_{P}\over
M_{\oplus}}\,\left({b_{P}\over b_\oplus}\right)^{-2}$,
where $a$ is the heliocentric distance in AU of the encounter, ``P'' and ``$\oplus$'' 
refer to the planet and to the
Earth. Saturn's tidal forces contribute $49\%$ of the actual separation D, while Jupiter 
contributes $11\%$. 
Since they are in a resonant region,  the
tidal features survive for 632 years, without
disappearing into the background. The size of the meteoroids prevente their depletion by
radiation pressure. 

\noindent
\textbf{Acknowledgments:}
Thanks are due to Dr.~Irene Sigismondi, Dr.~Renato Klippert  and Prof. William van Altena for
their valuable collaboration, and to Dr. Mark Gyssens and Dr. Rainer Arlt for their remarks. 
\begin{figure}
%\vspace{10cm}
%\hspace{0.5cm}
\hspace{0.5cm}
\resizebox{8cm}{4.5cm}{\includegraphics{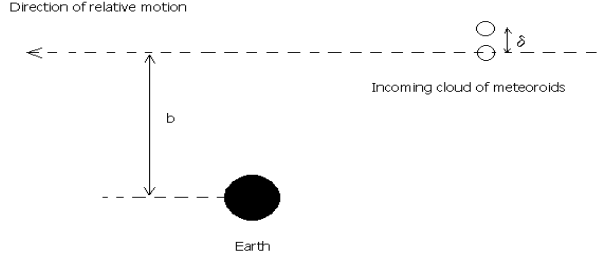}}
%\special{bmp: graincloud.bmp x=8cm y=4cm}
\caption[]{Geometry of the encounter in 1366 between the Earth and the
 cloud of meteoroids originating from the comet 55P/Tempel-Tuttle. 
The cloud is represented as a double body separated by $\delta$ upon which the tides operate. 
In 1366 the ``impact parameter" $b$ was $b_{\oplus}=0.025 AU$ [5].}
\label{fig.3}
\end{figure}

\noindent
\textbf{References:}
\noindent
[1] R. Arlt, P. Brown, WGN \textbf{27:6}, 267-285, (1999).
\noindent
[2] R. McNaught, D. Asher, WGN  \textbf{27:6}, 85-102, (1999).
\noindent
[3] D.J. Asher, M.E. Bailey, V.V. Emel'yanenko, MNRAS \textbf{304}, L53, (1999).
\noindent
[4] P. Jenniskens, Astron. Astrophys., \textbf{295}, 206, (1995).
\noindent
[5] D.K. Yeomans, K.Yau, P.Weissman, Icarus, \textbf{124}, 407, (1996).
\end{document}